\documentstyle[preprint,eqsecnum,aps]{revtex}

\begin{document}
\preprint{DOE/ER/40561-119-INT93-00-43}
\title{Light-Front Ensemble Projector Monte Carlo}
\author{M. Burkardt}
\address{
Institute for Nuclear Theory\\
Henderson Hall, HN-12\\
University of Washington\\
Seattle, WA 98195}
\date{\today}
\maketitle
\begin{abstract}
A new method to perform numerical simulations of light-front
Hamiltonians formulated on transverse lattices is introduced.
The method is based on a DLCQ formulation for the
(continuous) longitudinal directions. The hopping term
in the transverse direction introduces couplings between
fields defined on neighboring $1+1$-dimensional sheets.
Within each sheet, the light-cone imaginary time evolution operator
is calculated numerically with high precision
using DLCQ. The coupling
between neighboring sheets is taken into account using
an initial value random walk
algorithm based on the ensemble projector
Monte Carlo technique and a checkerboard decomposition for
the time evolution operator. The structure functions of
$\lambda \phi^4$ theory in $2+1$ dimensions are studied
as a trial application. The calculations
are performed with up to 64 transverse lattice sites.
No Tamm-Dancoff truncations are necessary.
\end{abstract}

\narrowtext

\section{Introduction}
\label{sec:intro}
Euclidean lattice $QCD$ allows one to calculate ground state
properties of hadrons,
but real time response functions are somewhere between difficult
and impossible. On the other hand, deep inelastic scattering gives
us information about correlation functions on
(or exceedingly close to) the light-cone. Light-front quantization
seems to be a promising tool to describe the immense wealth
of experimental information about structure functions for a
variety of reasons: (1) correlation functions along the
light-cone become ``static'' (i.e. equal $x^+=(x^0+x^3)/\sqrt{2}$)
observables in this approach. (2) structure functions are easy
to evaluate from the light-front wavefunctions. (3) these structure
functions are easily interpreted as light-front momentum densities.
However, before one can apply the light-front formalism to $QCD$
and other field theories, one has to remove the divergencies first
(i.e. regularize and renormalize).
One interesting idea in this direction is called the
``transverse lattice'' \cite{bardeen,paul,pauldoubl}.
Instead of discretizing all four space-time directions (like in
Euclidean lattice $QCD$) or the three space directions
(like in the Hamiltonian formulation of lattice $QCD$) one
discretizes only the two transverse ($x^1$ and $x^2$) directions
while leaving the longitudinal directions ($x^0$ and $x^3$)
continuous. On the one hand the transverse lattice thus provides
a gauge invariant UV regularization scheme and on the other hand
it is still possible to perform canonical light-front quantization
--- making it a promising approach towards performing
non-perturbative calculations of deep inelastic structure functions.

There remains the question what one should do with the
still continuous (i.e. infinitely many degrees of freedom)
longitudinal directions. First one may be tempted to
discretize the $x^-=(x^0-x^3)/\sqrt{2}$ (the light-cone space-)
direction as well. However, such an approach is suffering from
a fundamental difficulty: the longitudinal momentum is not
conserved on a longitudinal lattice. Due to Bragg reflections
it is only conserved modulo $2p^+_{max} = 2\pi/ a_L$, where
$a_L$ is the longitudinal lattice spacing.
Normally (i.e. in normal coordinates) this is not a problem
because the minimum of the kinetic energy occurs at
${\vec P} ={\vec 0}$.
However, the light-cone energy decreases with
increasing momentum ($P^- = M^2/2P^+$, in the continuum)
i.e. a minimum is reached for
$P^+ = \infty$! Of course on a lattice the momentum cannot
become infinite but still the minimum of the kinetic term
occurs around (depending on the precise form of the lattice
action) $p^+ = p^+_{max}/2 = \pi/2a_L$. Since the total momentum is
not conserved this implies that particles tend to
accumulate near this minimum. However, with such a large
momentum the particles can resolve the granular structure of the
lattice and no meaningful continuum limit will be obtained.
Note that a similar pathology would occur in
an unconstrained (total momentum allowed to vary) variational
calculation of the light-front energy of a hadron in the continuum.
It is conceivable that adding a Lagrange multiplier proportional
to the total light-front momentum to the lattice action
cures the problem
(in the continuum this amounts to minimizing
$\tilde{P}^- = P^- + \lambda P^+$ instead of $P^-$).
 However, this idea will not be
pursued here any further.

Instead, I found it more useful to work in momentum space
as far as the longitudinal direction is concerned
because this allows one easily to maintain longitudinal
momentum conservation --- a crucial necessity for light-front
calculations as we have seen above. One momentum space technique
which has been widely applied to light-front
quantized $1+1$-dimensional field theories
is discrete light-cone quantization
(DLCQ) \cite{dlcq1}. There one puts the system into a
longitudinal box and imposes periodic or antiperiodic boundary
conditions $\phi(x^-+L) = \pm \phi(x^-)$.
The momenta thus become discrete and solving the
equations of motion for a fixed value of $P^+$ has been
reduced to diagonalizing a finite matrix (note that all
light-cone momenta are positive and thus there is only a
finite number of states associated with a given value of $P^+$.
The longitudinal continuum limit is achieved by making
$P^+$ sufficiently large. In $1+1$ dimensions this technique
was very useful and effective \cite{dlcq2,deu}
For example, it has been used to demonstrate the existence of a
nucleon-nucleon bound state in $QCD_{1+1}$ with $SU(2)$-color
and $SU(2)$-flavor in a calculation on the level of quarks ---
despite the small binding energy ($\approx$ 1\%) of this
``deuteron'' \cite{deu}.
One mayor obstacle for applications of DLCQ to $2+1$ and $3+1$
dimensional field theories has been the exponential growth
of the number of basis states with the number of transverse
degrees of freedom. For example, a DLCQ calculation for a
scalar field (with antiperiodic boundary conditions
in the $x^-$ direction)
on a (rather modest) $4 \times 4$ transverse lattice
with a longitudinal momentum $P^+ = {15 \over 2}$ has
already a basis size of $779 022 208$. A more reasonable
$8 \times 8$ transverse lattice with the same $P^+$ requires
a basis of $6.27\cdot 10^{15}$ states!
These astronomical numbers clearly
demonstrate that any direct matrix diagonalization approach
or even a Lanczos type algorithm is doomed  to fail because
one is not even able to store the wavefunction in any available
computer.
Another numerical method for studying light-front
Hamiltonians is the light-front Tamm-Dancoff approach \cite{lftd}
where one imposes severe truncations
of the Fock space. Sometimes this method is combined with DLCQ,
i.e. one formulates the light-front Hamiltonian in the DLCQ basis
but restricts ({\it ad hoc}) the Fock space to few particle states.
However, it is not clear to what extend this
truncation modifies the dynamics and whether effective
light-front Hamiltonians can be constructed by systematically
eliminating higher Fock states.
Because of these difficulties, all numerical studies
of light-front Hamiltonians have been restricted to
$1+1$ dimensions and/or severe Tamm-Dancoff truncations and/or
perturbation theory.

In this work Monte Carlo techniques will be exploited
to obtain approximate ground state energies and
structure functions for DLCQ problems in $2+1$ and $3+1$
dimensions on a transverse lattice without any constraints
on the Fock space other than those resulting form the
discreteness of the momenta. Note that the algorithm which I will
present requires that the interaction in the transverse
direction is local (nearest neighbor interaction at most).
That is why the theory will be formulated on a transverse
coordinate space lattice. The algorithm would not work with
a momentum space lattice. To avoid obscuring the
Monte Carlo algorithm with other difficulties the technique will
be explained for a real scalar field with $\phi^4$ coupling
in $2+1$ dimensions, which is one of the most simple field theories
one can formulate on a transverse lattice. The main reason to chose
$\phi^4_{2+1}$ for illustrating this new method is that $QCD$ is
too complicated for a ``first study case'' and for demonstration
purposes. Other theories one might think of,
like $QED_{3+1}$ of $\phi^4_{3+1}$ are not asymptotically free,
i.e. there is no Bjorken scaling for deep inelastic
structure functions.
However, it should be emphasized that the technique is {\it a priori}
applicable to any DLCQ Hamiltonian which is formulated on a
transverse lattice, provided the interactions satisfy locality
in the transverse direction.

The paper is organized as follows. First the DLCQ Hamiltonian
for $\phi^4$ in $2+1$ dimensions on a transverse lattice
will be constructed. In section \ref{sec:mcalg}, the infinitesimal
light-front time evolution operator will be approximated
using a checkerboard decomposition and a path integral
in the Fock space of DLCQ will be
used to project out the ground state for given
quantum numbers. Finally, in section IV, the path integrals will be
evaluated using an initial value random walk algorithm
based on a variation of
the ensemble projector Monte Carlo method.

\section{$\phi^4_{2+1}$ on a transverse lattice}
\label{sec:dlcqham}
The Minkowsky action for the $\phi^4_{2+1}$ model, in the continuum,
reads
\begin{equation}
A_{cont.} = \int d^3x \frac{1}{2} \partial_\mu \phi \partial^\mu \phi
- \frac{m^2}{2} \phi^2 - \frac{\lambda}{4!}\phi^4.
\end{equation}
After discretization in the transverse direction one thus obtains
\begin{eqnarray}
A_{\perp latt.} = a \int dx^+dx^- \sum_n
\Bigg[ \partial_+ &\phi_n &\partial_- \phi_n
- \frac{m^2}{2} \phi_n^2 - \frac{\lambda}{4!} \phi_n^4
\nonumber\\
&-&\left. \frac{ \left( \phi_{n+1} - \phi_n \right)^2}
{2a^2}
\right].
\end{eqnarray}
Upon rescaling $\varphi_n = \sqrt{a}\phi_n$ the interpretation
of the transverse lattice action as the action of a
``multiflavor'' field theory ($n$ being the ``flavor'' index in
this interpretation and $\varphi_n$ being a canonical
field defined in $1+1$ dimensions) becomes evident
\begin{eqnarray}
A_{\perp latt.} = \int dx^+dx^- \sum_n
\Bigg[ \partial_+ &\varphi_n &\partial_- \varphi_n
- \frac{m^2}{2} \varphi_n^2 - \frac{\lambda}{a4!} \varphi_n^4
\nonumber\\
&-&\left. \frac{ \left( \varphi_{n+1} - \varphi_n \right)^2} {2a^2}
\right].
\label{eq:action}
\end{eqnarray}
Light-front quantization of (\ref{eq:action})
is standard \cite{hari}: first one puts the system into an
$x^-$-box of length $L$ with antiperiodic boundary conditions
(the associated zero-mode effects and implications for the
renormalization are discussed in Ref.\cite{mbsg}). The canonical
commutation relations
\begin{equation}
\left[ \partial_-\varphi_n(x), \varphi_m(y) \right]_{x^+=y^+}
= -\frac{i}{2} \delta_{nm} \delta(x^--y^-)
\end{equation}
as well as the antiperiodic boundary condition are satisfied for
\begin{equation}
\varphi_n(x^-) = \frac{1}{\sqrt{4\pi}} \sum_{k=1}^\infty
\frac{\left[ a_n(k)e^{-ip_k^+x^-}  +
a_n^\dagger(k)e^{ip_k^+x^-}\right]}{\sqrt{k - \frac{1}{2}}}
\label{eq:phiexp}
\end{equation}
where
\begin{equation}
p_k^+ = \frac{2\pi}{L} \left( k - \frac{1}{2} \right)
\label{eq:pk}
\end{equation}
and the $a_n(k)$ satisfy the usual commutation relations
\begin{equation}
\left[ a_n(k), a_m^\dagger(q)\right] = \delta_{nm} \delta_{qk}.
\end{equation}
Finally one obtains for the light-front momentum operator
\begin{eqnarray}
P^+ &=& \sum_n \int dx^- :\partial^+ \varphi_n \partial ^+ \varphi_n:
\nonumber\\
&=& \frac{2\pi}{L} \sum_n \sum_{k=1}^\infty a_n^\dagger(k)a_n(k)
\left( k - \frac{1}{2} \right)
\end{eqnarray}
and for the light-front energy
\begin{equation}
P^- = \frac{L}{2\pi} \sum_n \left( T_n + V_n + V_{n,n+1} \right)
\end{equation}
where
\begin{equation}
T_n= \frac{m^2}{2} \sum_{k=1}^\infty \frac{a_n^\dagger (k) a_n(k)}
{k-\frac{1}{2}}
\end{equation}
is the usual light-front kinetic energy on each site and
\begin{equation}
V_n = \frac{\lambda \delta_{P_f,P_i}}{4\pi a 4!}
\sum_{k_1,k_2,k_3,k_4=1}^\infty
\frac{:\left(a_n^\dagger(k_1) + a_n(k_1) \right)}
{\sqrt{k_1-\frac{1}{2}}}
\frac{\left(a_n^\dagger (k_2) + a_n(k_2) \right)}
{\sqrt{k_2-\frac{1}{2}}}
\frac{\left(a_n^\dagger (k_3) + a_n(k_3) \right)}
{\sqrt{k_3-\frac{1}{2}}}
\frac{\left(a_n^\dagger (k_4) + a_n(k_4) \right):}
{\sqrt{k_4-\frac{1}{2}}}
\end{equation}
is, up to the factor $1/a$ the self interaction for
$\varphi^4_{1+1}$. $\delta_{P_f,P_i}$ is a momentum
conserving Kronecker $\delta$. Neighboring sites
are coupled through the hopping term
\begin{equation}
V_{n,n+1}= \frac{1}{2a^2} \sum_{k=1}^\infty
\frac{\left( a_{n+1}^\dagger (k)-a_{n}^\dagger (k)\right)
\Bigl( a_{n+1} (k) -a_{n}(k) \Bigr)}
{k - \frac{1}{2}}
{}.
\end{equation}
Note that, as in all DLCQ problems,
the length of the box factorizes completely.
Thus we will in the following work with the rescaled operators
\begin{equation}
K=P^+ L/2\pi
\label{eq:kdlcq}
\end{equation}
 and
\begin{equation}
H=P^-2\pi/L.
\label{eq:hdlcq}
\end{equation}
At least in principle one could now proceed as follows: for fixed $K$
($K$ and $H$ commute) one diagonalizes $H$, yielding the eigenvalue
$E_i$ and thus the invariant masses of the physical states
$M_i^2 = 2 K E_i$.
Physical observables like structure functions are obtained
by calculating the appropriate matrix element in these states.
The continuum limit is reached by extrapolating
to $K \rightarrow \infty$. For one or two sites this can be
easily done. For four sites
this is also still possible. However, beyond about $8$ sites
direct diagonalization methods (including the Lanczos algorithm)
soon become useless due to the exponential growth of the required
basis size. What is needed at this point is a Monte Carlo method that
allows one to find the ground state (for given value of $K$)
of the DLCQ Hamiltonian $H$ more efficiently.
Developing and testing such an
algorithm will be the main subject in the rest of this paper.

Before we come to the Monte Carlo algorithm I should add a
few comments. First, although I have derived the transverse
lattice DLCQ Hamiltonian only for $\phi^4_{2+1}$, similar
expressions can be derived for $3+1$ dimensions and/or other
field theories \cite{bardeen,paul,pauldoubl}.
Many aspects are completely general. For example,
for many field theories (including $QCD_{3+1}$ \cite{bardeen})
the transverse ``hopping'' is provided by a nearest
neighbor interactions. This is crucial for the algorithm which
I will present in this paper because
(i) one can easily approximate the infinitesimal light-front
time evolution operator as will be explained in section \ref{sec:mcalg}
(ii) it is possible to ``locally update'' the states in a
random walk algorithm which thus provides a computational advantage
on large transverse lattices.
Second, I should discuss renormalization
at this point. Besides the tadpoles (which are zero in light-front
quantization \cite{mbsg,paulsg}) there is only one divergent diagram
in $\phi^4_{2+1}$, namely the setting sun diagram (Fig.\ref{oyster}).
This diagram leads to a divergent self mass contribution while
the associated wave function renormalization is finite. Therefore,
for $\phi^4_{2+1}$ it will be sufficient to add an appropriate
mass counterterm to render the theory finite in the continuum limit.
Here one has to be a little careful in DLCQ because
the longitudinal momentum of the incoming line in (Fig.\ref{oyster})
effectively determines which higher Fock states are allowed
in the intermediate state. Hence the self mass, and thus the
required mass counterterm, for (Fig.\ref{oyster}) depend on the
longitudinal momentum (of course, for large longitudinal momenta,
this dependence asymptotically disappears). For the renormalization
this leaves at least two options: one possibility is
to calculate how the infinite part of the self energy depends on the
longitudinal momentum by evaluating (Fig.\ref{oyster}) within
DLCQ. Or one allows the bare mass to depend on the longitudinal
momentum in such a way that the physical mass of the lightest
state is independent of the longitudinal momentum. This is possible
if one uses the following ``sequential'' procedure:
First one solves the DLCQ Hamiltonian for $K=1/2$ (which
is trivial) and adjusting the bare mass for $K=1/2$ until
one satisfies the renormalization condition. Then
one repeats the same procedure for $K=3/2$ while keeping the bare
mass for $K=1/2$ partons fixed and so on.
In the continuum limit
(large longitudinal momenta)
both methods should be equivalent. However, I preferred to use the second
method (sequential renormalization) because it seemed to converge
faster numerically in $K$. Furthermore, because one must verify
numerical convergence in $K$, it is anyway necessary to repeat
the calculation for several $K$. In addition, the computer time
spent usually grows rapidly with $K$, i.e. the numerical effort
associated with additional calculations
for small values of $K$ in the sequential renormalization
is negligible. Another advantage of renormalizing such that
$M_{phys}$ is independent of $K$ is that multiparticle
thresholds in two point functions appear at the correct
energy values relative to the single particle pole.
This issue has often been neglected in DLCQ calculations.
Furthermore, while the first method works
only for $\phi^4_{2+1}$ because there is
only one divergent self-energy diagram in this theory,
the sequential method is non-perturbative and can thus be applied
to the DLCQ Hamiltonian for any field theory.

The coupling constant for $\phi^4_{2+1}$ receives only a
finite renormalization. Nevertheless, in principle one should
always fix the bare coupling by imposing another renormalization
condition. Since the spectrum of $\phi^4_{2+1}$ consists only
of the fundamental particle and its scattering states, and since
the mass of the fundamental particle has already been used up
to fix the bare mass term, one has to use physical observables
other than the mass spectrum in this case. However, since this is a
special feature for $\phi^4_{2+1}$ I decided not to renormalize
the coupling at this point. In the following, all results will
be quoted with the bare coupling constant for which they were
calculated.
\section{The Monte Carlo Procedure}
\label{sec:mcalg}
In this section, we will use the fact that repeated
application of $\exp (-\varepsilon H)$, where $H$ is the
DLCQ Hamiltonian (\ref{eq:hdlcq}), on any state
$|\psi (K)>$ with given light-cone momentum $K$ (\ref{eq:kdlcq})
gives the ground state $|\psi_0 (K)>$ for this particular $K$ ---
of course provided $|\psi (K)>$ is not orthogonal to the
ground state $|\psi_0 (K)>$. The light-front imaginary time
evolution operator $\exp (-\varepsilon H)$
can be approximated using the Trotter formula \cite{pmc1}
\begin{eqnarray}
e^{-\varepsilon (H_a+H_b)}&=&
e^{-\varepsilon H_a/2}e^{-\varepsilon H_b}e^{-\varepsilon H_a/2}+
\nonumber\\
& & \frac{\varepsilon^3}{24}\left\{
\left[ H_a, [H_b,H_a] \right] - 2 \left[ H_b, [H_a,H_b] \right]
\right\} + ...
\nonumber\\
&=&e^{-\varepsilon H_a/2}e^{-\varepsilon H_b}e^{-\varepsilon H_a/2}
+ {\cal O}(\varepsilon^3)
\label{eq:exphab}
\end{eqnarray}
where the following choice is particularly useful
\begin{eqnarray}
H_a &=& \frac{1}{2} \sum_n \left( T_n+V_n \right)
+ \sum_{n\ odd}V_{n,n+1}
\nonumber\\
H_b &=& \frac{1}{2} \sum_n \left( T_n+V_n \right)
+ \sum_{n\ even}V_{n,n+1}.
\end{eqnarray}
This choice is motivated because $H_a$ and $H_b$
individually separate into sublattices which contain only
2 sites each
\begin{eqnarray}
H_a&=& \left[ \frac{1}{2}\left(T_1+V_1+T_2+V_2\right) +V_{1,2}\right]
+
\nonumber\\
& & \left[ \frac{1}{2}\left(T_3+V_3+T_4+V_4\right) +V_{3,4}\right]
+...
\nonumber\\
&=& H_{1,2} + H_{3,4} +...
\nonumber\\
H_b&=& \left[ \frac{1}{2}\left(T_2+V_2+T_3+V_3\right) +V_{2,3}\right]
+
\nonumber\\
& & \left[ \frac{1}{2}\left(T_4+V_4+T_5+V_5\right) +V_{4,5}\right]
+...
\nonumber\\
&=& H_{2,3} + H_{4,5} +...
\end{eqnarray}
i.e. both $H_a$ and $H_b$ are sums of commuting operators and
can be easily diagonalized and
exponentiated numerically! This task is facilitated even more by
the fact that $H_{1,2}$, $H_{2,3}$ etc. each satisfy longitudinal
momentum conservation, i.e. $H_{1,2}$, $H_{2,3}$ etc. are all
block diagonal in the DLCQ basis, where the blocks
are labeled by the sum of momenta on sites 1 and 2, etc. !

Once one has constructed $\exp (-\varepsilon H)$ one
can proceed to evaluate physical observables in the ground state,
using
\begin{equation}
E_0(K) = \lim_{N \rightarrow \infty}
\frac{ <\psi_f(K)| H \left( e^{-\varepsilon H}\right)^N |\psi_i(K)>}
{ <\psi_f(K)| \left( e^{-\varepsilon H}\right)^N |\psi_i(K)>}
\label{eq:hpath}
\end{equation}
for the ground state energy and
\begin{eqnarray}
&&\frac{<\psi_0(K)| \hat{O} |\psi_0(K)>}{<\psi_0(K)|\psi_0(K)>}
= \nonumber\\
&&\lim_{M,N\rightarrow \infty}
\frac{ <\psi_f(K)|\left( e^{-\varepsilon H}\right)^M \hat{O}
\left( e^{-\varepsilon H}\right)^N  |\psi_i(K)>}
{ <\psi_f(K)| \left( e^{-\varepsilon H}\right)^{M+N} |\psi_i(K)>}
\label{eq:opath}
\end{eqnarray}
for any other observable, provided $<\psi_0(K)|\psi_i(K)> \neq 0$
and $<\psi_f(K)|\psi_0(K)> \neq 0$.
The results thus obtained still depend on $\varepsilon$ because
$\exp (-\varepsilon H)$ has been approximated only up to
${\cal O}(\varepsilon^3)$. The $\varepsilon \rightarrow 0$ limit
can be obtained by extrapolation.

The above products are evaluated as follows. First one inserts a
complete set of states at each $\varepsilon$-step, e.g.
\widetext
\begin{eqnarray}
<\psi_f(K)| \left( e^{-\varepsilon H}\right)^N |\psi_i(K)>=
\sum_{{\vec i}_l\ {\vec k}_l}&&
<\psi_f(K)| e^{-\frac{\varepsilon}{2} H_a}
|{\vec i}_{3N+1}\ {\vec k}_{3N+1}>
<{\vec i}_{3N+1}\ {\vec k}_{3N+1}|
e^{-\varepsilon H_b}
|{\vec i}_{3N}\ {\vec k}_{3N}>
\nonumber\\
&&\times
<{\vec i}_{3N}\ {\vec k}_{3N}|\ .\ .\ .\
e^{-\frac{\varepsilon}{2} H_a}|{\vec i}_1\ {\vec k}_1>
<{\vec i}_1\ {\vec k}_1|\psi_i(K)>,
\label{eq:evolv}
\end{eqnarray}
\narrowtext
where the $|{\vec i}_l\ {\vec k}_l>$
($l=1,...,3N+1$) are a direct product of
complete sets of states at each site. The vector notation is used
to label these states where the n-th component of ${\vec k}$
refers to the longitudinal momentum on the n-th site and the
n-th component of ${\vec i}$ refers to the internal excitation
level on this n-th site with a given momentum\footnote{After all
there is still a $1+1$ dimensional field theory attached to each site,
i.e. for each site-momentum many different states are possible.}
For example for a lattice with 4 sites a randomly picked
state could look like this:
\begin{eqnarray}
{\vec k} &= \left( 0, 1, \frac{5}{2}, 3\right)\nonumber\\
{\vec i} &= \left( 1, 1, 3, 2\right)
\end{eqnarray}
which would mean that site one carries $0$ longitudinal momentum,
site two $1$ unit, etc. while the internal excitation numbers are
$1$ (for zero momentum there is only one state --- the vacuum),
$1$, $3$ and $2$ respectively.
Of course one can chose many other ways to enumerate the basis
but I found the above method the easiest to implement
in a computer code.

In this work, a free Fock space basis has been selected to
represent the internal excitations at
each site because
structure functions are diagonal only in a free Fock space basis.
However, many other choices are conceivable as well.
In fact, in many cases it may be more efficient numerically
to chose some kind of interacting basis on the sites instead of
a free basis. Particularly if one is more interested in energy
eigenvalues or observables other than structure functions.

Several Monte Carlo techniques are available to perform the
summations in Eq.(\ref{eq:evolv}). I have tried the projector
Monte Carlo method \cite{pmc1} as well as the
ensemble projector Monte Carlo method \cite{epmc1}.
In the projector Monte Carlo method, in each sweep one starts
from a state $|{\vec i}_1\ {\vec k}_1>$ which is randomly
picked with probability $<{\vec i}_1\ {\vec k}_1|\psi_i(K)>$.
In the next step one picks a state $|{\vec i}_2\ {\vec k}_2>$
with probability
$W({\vec i}_1, {\vec k}_1 \rightarrow {\vec i}_2, {\vec k}_2)$.
The probabilities chosen in this work are
\footnote{In principle, there is little restriction in
the choice of the ``probabilities''. However, I found
Eq.(\ref{eq:prob}) useful numerically.}
\begin{equation}
W({\vec i}_l, {\vec k}_l \rightarrow {\vec i}_{l+1}, {\vec k}_{l+1})
= \frac{|<{\vec i}_{l+1}\ {\vec k}_{l+1}|U|{\vec i}_{l}\
{\vec k}_{l}>|}
{\sum_{ {\vec i}\ {\vec k}}
|<{\vec i}\ {\vec k}|U|{\vec i}_{l}\ {\vec k}_{l}>|}
\label{eq:prob}
\end{equation}
where $U=\exp (- \frac{\varepsilon}{2} H_a )$ for $l=3N$ or $l=3N+1$
and $U=\exp (-\varepsilon H_b)$ for $l=3N+2$
, $N=0,1,2,..\ $.
\footnote{Note that two
adjacent steps with $U=\exp (- \frac{\varepsilon}{2} H_a )$
can be combined into one step with $U=\exp (- \varepsilon H_a )$.}
Since $U$ factorizes into two-site sublattices, so does the
transition probability $W$. For example, suppose (on a lattice with
$4$ sites)
\begin{eqnarray}
{\vec k}_l =& (k^1_l, k^2_l, k^3_l, k^4_l)
\nonumber\\
{\vec i}_l =& (i^1_l, i^2_l, i^3_l, i^4_l)
\end{eqnarray}
and suppose $U=\exp (- \frac{\varepsilon}{2} H_a )$. Then one first
selects $i_{l+1}^1, k_{l+1}^1, i_{l+1}^2, k_{l+1}^2$
\footnote{Note that momentum conservation requires
$k^1_{l+1} + k^2_{l+1} = k^1_l + k^2_l$ and thus restricts the
possible choices.}
with probability proportional to
\begin{equation}
\left|<i_{l+1}^1, k_{l+1}^1, i_{l+1}^2, k_{l+1}^2|
e^{- \frac{\varepsilon}{2} H_{1,2} }
|i_{l}^1, k_{l}^1, i_{l}^2, k_{l}^2>\right|.
\end{equation}
The actual selection can be done with a Metropolis algorithm.
Then one applies the same procedure to sites 3 and 4.
Having thus ``updated'' the entire lattice
one proceeds to the next $\varepsilon$-step where
$U=\exp (-\varepsilon H_b )$. This time it is sites 2 and 3
as well as sites 4 and 1 which interact with another.
And so on until one reaches $<{\vec i}_{3N+1}\ {\vec k}_{3N+1}|$.
For such a ``path'' one then computes the product
\begin{equation}
S = S_{f,3N+1} \times S_{3N+1,3N} \times ... \times S_{2,1},
\label{eq:stimess}
\end{equation}
where
\begin{equation}
S_{f,3N+1} = <\psi_f(K) |e^{-\frac{\varepsilon}{2} H_a}|
{\vec i}_{3N+1}\ {\vec k}_{3N+1}>
\end{equation}
and the ``scores'' at each step are the actual transition
amplitudes (the matrix elements of $U$)
divided by the ``probabilities''
\begin{equation}
S_{l+1,l} = \frac{
<{\vec i}_{l+1}\ {\vec k}_{l+1} | U | {\vec i}_l\ {\vec k}_l>}
{W({\vec i}_l, {\vec k}_l\rightarrow {\vec i}_{l+1},
{\vec k}_{l+1})}.
\end{equation}
These products of scores (\ref{eq:stimess})
are then averaged over many sweeps.
Similarly one computes the numerators in Eqs.(\ref{eq:hpath})
(\ref{eq:opath}). For example, if one wants to calculate
a structure function, one averages over the product of scores
times the structure function measured somewhere near the middle
of the path and divides the result by the average of the
product of scores.\footnote{Here it becomes clear why a
diagonal representation is preferable for an easy computation
of structure functions.}

The disadvantage of the projector Monte Carlo method
is that once a small ``score'' occurs along  a path,
the whole product for the path will contribute negligibly to the
sum of paths, i.e. the path integral will be dominated by a few
paths which do not contain any small element and the relative
statistical fluctuations in the path thus increase
linearly with the number of steps in the path
\cite{epmc1,epmc2}.
Since on the other hand one is interested in making the path
as long as possible in order to project onto the ground state
while keeping $\varepsilon$ small to avoid systematic errors, the
projector Monte Carlo method turns out to be rather inefficient.
A variation of this method, the ensemble projector Monte Carlo
method suffers less from this problem. There one starts form
an ensemble of states at step 1. The probabilities and scores are
calculated for each member of the ensemble in the same way
as for the projector Monte Carlo method.
However, after each $\varepsilon$-step, when one calculates
the scores, one replicates and deletes members of the ensemble
by the following rule: first one calculates the average score
$\bar{S}$ (ensemble average). Then one replicates each state $\nu$
in the ensemble with multiplicity
\begin{equation}
n_\nu = Int \left[ \frac{|S_\nu |}{\bar{S}}\
+\  'random\ number \in (0;1)' \right],
\label{eq:multi}
\end{equation}
where $S_\nu$ is the score for the transition to state $\nu$.
This way, any path where a very small score occurs will most
likely be eliminated (and no further computer time will be
wasted on this path) while important paths branch out and
contribute with multiple weight. Note that the size of the
population is not fixed because the states are replicated
with a multiplicity that depends on a random number.
In general, the population thus fluctuates. Sometimes,
the population grows or shrinks slowly. In order to keep
the population stable on the average one can counterbalance
the growing/shrinking by choosing $\bar{S}$ in Eq.(\ref{eq:multi})
a little larger/smaller than the average score.

For the path integral one only has to evaluate the sign of the
scores
\begin{equation}
\tilde{S} = S_{f,3N+1} \times
sign\left( S_{3N+1,3N} \times ... \times S_{2,1} \right).
\label{eq:sprod}
\end{equation}
The absolute values of the scores along the path are already
taken into account since they determined the multiplicities
in the random walk (\ref{eq:multi}).
Observables are calculated similarly as in the projector Monte
Carlo Method except that one now averages over both
the ensemble and the sweeps.
For the efficiency of the algorithm, it is important to
start the random walk with a good initial guess for the ground
state $|\psi_i(K)>$. For many light-front Hamiltonians the exact ground
state wavefunction has a sizeable overlap with the valence configuration.
As a first try, it may thus be useful to make a valence ansatz
for $|\psi_i(K)>$ and $|\psi_f(K)>$. This will also be the choice in this
work. For more complicated theories one can also try to improve
the valence ansatz for $|\psi_i(K)>$ by including higher Fock
components perturbatively.

\section{Numerical results}
The Lagrangian for $\phi^4_{2+1}$ contains two dimensionful
parameters: the bare mass $m_0$ and the coupling constant
$\lambda$ (which also carries dimension of mass) and hence
an arbitrary scale $\Lambda$, specifying the units in which
these dimensionful parameters are measured.
In the following, the scale will be fixed by demanding that
$M_{phys} = \Lambda$ for the lightest physical particle. All
other dimensionful quantities (e.g. $\lambda$ or $a^{-1}$)
will be measured in these units. After fixing the physical mass
scale, which determines the bare mass,
there is still (in the continuum limit) one
dimensionless parameter left: $\lambda / 4\pi \Lambda$. In
my numerical work I have considered two extreme cases:
$ \lambda / 4\pi \Lambda =1$ and $ \lambda / 4\pi \Lambda =10$
corresponding to intermediate and strong coupling respectively.
The case $ \lambda / 4\pi \Lambda =1$ turned out to be rather
boring because the structure function of the lightest
physical particle is strongly dominated by a ``bare''
excitation $|\psi_0(K)> \approx \sum_n a^\dagger_n(K) |0>$.
Deviations from this pointlike structure
can be well approximated by summing
a chain of ``setting suns''(Fig.\ref{oyster}).
Although this result was reproduced in the Monte Carlo calculations,
it will not be discussed here any further because the real
strength of the Monte Carlo method lies more in the
nonperturbative regime.
$ \lambda / 4\pi \Lambda =10$ will thus be chosen throughout the
rest of this paper.

The numerical calculations were done as follows: after selecting
the coupling constant ($ \lambda / 4\pi \Lambda =10$) and
choosing a transverse lattice (characterized by the spacing
$a\Lambda$ and by the number of sites) and after selecting
a value for the ``damping parameter'' $\varepsilon$, the
bare masses were determined by imposing the renormalization
condition $M_{phys} / \Lambda =1$. Within the DLCQ
formalism used here this yields bare masses which show
some dependence on the longitudinal momentum.
For $p=1/2$ there is no interaction and
$M_{phys}(\frac{1}{2}) = \Lambda $ implies
$m_0^2(\frac{1}{2}) = \Lambda^2$.
In the next step the bare mass $m_0^2(\frac{3}{2})$ for $p=3/2$
is determined by requiring $M_{phys}(\frac{3}{2}) = \Lambda $
(keeping $m_0^2(\frac{1}{2}$) fixed). Then $m_0^2(\frac{5}{2})$
while keeping $m_0^2(\frac{3}{2})$ and $m_0^2(\frac{1}{2})$
fixed and so on, up to $p=\frac{15}{2}$ --- the largest
momentum used in this work.
In this fine-tuning process, at every longitudinal
momentum, the physical mass of the lightest
particle was determined
using the ensemble projector Monte Carlo technique with an
ensemble size of $500$ states and with $10000$ $\varepsilon$-steps.
For $\psi_i$ and $\psi_f$ a plane wave
(zero transverse momentum) of bare ground state
``mesons'' was used
\begin{equation}
|\psi_i(K)> = |\psi_f(K)> = \sum_n a^\dagger_n(K) |0>.
\label{eq:psiif}
\end{equation}
This choice (which corresponds to the valence approximation in $\phi^4$)
was motivated by the fact that the physical ground state particle
in $\phi^4_{2+1}$ can be interpreted as a dressed single particle state.

After 15 ``thermalization steps'' the energy was sampled
every 5th $\varepsilon$-step \footnote{Here each application
of $\exp (-\varepsilon H_a)$ or $\exp (-\varepsilon H_b)$ is
counted as one $\varepsilon$-step.}
(to insure statistically
independent sampling) by separately taking the ensemble
average of the numerator and the denominator in Eq.(\ref{eq:hpath}).
The physical mass is obtained from
\begin{equation}
M_{phys}^2 = 2K \bar{E},
\label{eq:mphys}
\end{equation}
where $\bar{E}$ is the average over all energy measurements
($10000/5=2000$ in the above procedure). Typical (statistical)
errors with these parameters where of the
order of $1-3\%$ for $\bar{E}$  and thus
also for the bare masses. Whenever $M_{phys}$, evaluated using
Eq.(\ref{eq:mphys}) deviated significantly from $\Lambda$
(the renormalization condition) the bare masses where adjusted
accordingly.

This procedure was repeated for $\varepsilon \Lambda = .3,\
.15,\ .075$ and for lattices with
$N_{sites}=4,\ 8,\ 16,\ 32$ and $64$
transverse sites and for transverse spacings
$a\Lambda = 1,\ \frac{1}{2},\ \frac{1}{4}$.
It turned out that the bare masses depend only very weakly on
$\varepsilon$ and $N_{sites}$ which made the tuning rather easy.
The tuning was facilitated even more by the fact that the
path integral can still be summed up numerically exactly
(without Monte Carlo) for lattices with four sites (and
$p\leq \frac{15}{2}$). Furthermore, one can estimate the difference
between 4-site lattices and larger lattices by perturbative
methods before one starts the nonperturbative tuning using the
Monte Carlo.
For $a=1/4$ and large $K$ the energy measurement described above
became too noisy due to the sign problem discussed at the end
of section \ref{sec:mcalg}. For these cases, I used a slightly
different algorithm: Instead of evolving the initial state
for $10000\ \varepsilon$-steps and sampling the energy every
fifth step, the initial state was evolved only for
$10\ \varepsilon$-steps. This procedure was then repeated 2000
times to obtain the same statistical sample size
for the energies as in the
first method. The advantage of the second procedure is the
following. Since only a few (small) matrix
elements of $\exp (-\varepsilon H)$ are negative, it is not
very likely to encounter a negative score in a given ``path''
(\ref{eq:sprod}) --- unless the path is very long.
With this scenario it is clear that procedure two (starting
over and over again from the same initial state) has much less
of a sign problem than procedure one (continued evolution).
However, typically it takes more
$\varepsilon$-steps to project onto the ground state
from $|\psi_i(K)>$ (\ref{eq:psiif})
than it takes to get uncorrelated energy measurements.
It thus depends on the
concrete example which of the two procedures is more efficient.

After completing the renormalization for a given set of
parameters ($N_{sites},\ a,\ \varepsilon,\ \lambda$) one can
proceed to evaluate physical observables.
At this point let me introduce the ``structure functions'' for
scalar fields. In analogy to definitions of parton distributions
in $QCD$ one can introduce
\begin{equation}
f(x) = x \int_{-\infty}^{\infty}
\frac{d\xi^-}{4\pi} e^{ix\xi^- p^+}
<\psi(p^+)| \phi(0) \phi(\xi^-) | \psi(p^+)>
\end{equation}
as the light-cone momentum density of elementary quanta
in the state $|\psi(p^+)>$\footnote{One can in fact imagine
gedanken experiments that would allow one to measure $f(x)$
but this point will not be discussed here any further.}.
The normalization is such that the momentum sum rule
reads $\int_0^1 dx\ xf(x) =1$. Upon discretizing the structure
function can be expressed as (\ref{eq:phiexp})(\ref{eq:pk})
\begin{equation}
f(x_p) = \frac{1}{K} \sum_{n=1}^{N_{sites}} <\psi(K)|
a_n^\dagger(p) a_n(p) |\psi(K)>
\label{eq:fdlcq}
\end{equation}
where $x_p=(p-\frac{1}{2})/K$, $p=1,...,K+\frac{1}{2}$.
This expression (\ref{eq:fdlcq}) is the form used in this work.
What should one expect $f(x)$ to look like for $\phi^4_{2+1}$?
First, for trivial kinematical reasons, $f(x)$ is nonzero only
for $x \in [0;1]$. Second, since the wavefunction renormalization
in $\phi^4_{2+1}$ is finite, there is a finite probability
to find the physical ``meson'' as a bare state. Thus, in the
continuum limit,  $f(x)$
should contain a $\delta$-function at $x=1$ with finite coefficient
(in the structure function plots in this work this point will
always be excluded because it would lie outside the chosen plotframe).
Besides the $\delta$-function one expects a continuum because the
bare state can always split into three ``partons''
(via the $\phi^4$-interaction) which can split again and so on. While
lowest order perturbation theory suggests a structure function that
is peaked around $x={\cal O}(\frac{1}{3})$, higher order effects
(multifragmentation)
will shift the maximum towards smaller values of $x$.

Structure functions were evaluated by averaging over $10000$
sweeps (except for $a=0.25$ where structure functions were
averaged over 20000 sweeps)
where again an ensemble size of $500$ was used. Each
sweep consisted of $30\ \varepsilon$-steps.
Also $\psi_i = \psi_f = $ Eq.(\ref{eq:psiif}) was used again.
The operator to measure the structure functions
was inserted after the first $15\ \varepsilon$-steps
for each member of the ensemble.\footnote{Note that the
structure functions in DLCQ are defined for a discrete set
of points only (8 points for $K=\frac{15}{2}$).
Hence ``measuring the structure function'' means
evaluating it at a few points (\ref{eq:fdlcq}).}
For the remaining steps of the random walk, a record is kept
of the result of this structure function measurement from
which each subsequent ensemble member has evolved \cite{epmc2}.
Finally, the result for the measurement of $f(x)$ in this
sweep is obtained by separately evaluating
the ensemble average of the numerator and the denominator
in Eq.(\ref{eq:opath}). This result is then averaged over
the sweeps. In order to investigate convergence
with respect to the number of $\varepsilon$-steps in the final state,
this number was kept
variable ($1-15$). Typical results for such a structure function
measurement, as a function of the number of $\varepsilon$-steps
before taking the overlap with $<\psi_f(K)|$, are shown
in Fig.\ref{plateau}. For all values of $\varepsilon$ used in this
work, the plateau sets in before $10\ \varepsilon$-steps, i.e.
the ground state expectation value of the structure function
can be extracted from these results.\footnote{Since $\psi_i = \psi_f$,
conclusions
about convergence in the final state can also be applied to the
convergence in the initial state as well.}
There are several reasons for this rapid convergence.
First, even for $\lambda/4\pi \Lambda = 10$,
the true ground state has a large overlap with the bare state
$|\psi_i>$ (\ref{eq:psiif}). Furthermore, any contamination of
$|\psi_i>$ with excited states is filtered out very efficiently
because higher states are suppressed by the square of their
masses
\begin{equation}
\exp (-\varepsilon H) = \sum_n |n><n| \exp ( -\varepsilon
\frac{ M_n^2}{2K}).
\end{equation}
The lowest excited state with the same quantum numbers as
$|\psi_i>$ is a scattering state consisting of $3$ ground
state mesons with a threshold at $M_1 = 3M_0$
(i.e. $M_1^2 = 9 M_0^2$ !). For example, after $15$ steps with
$\varepsilon =0.3$, $M_0^2=1$ and $K=\frac{15}{2}$
the ground state is enhanced by a factor of
$e^{-0.3}/e^{-2.7} \approx 11$ compared to the first
excited state. Since most of the ``contamination'' comes not
from the threshold itself but from many higher excited
states, the filtering process is even more
efficient than this numerical example illustrates.

Since the structure functions turned out to have converged already
after $10\ \varepsilon$-steps in the final state,
\footnote{Note that this means that $10\ \varepsilon$-steps
would also have been sufficient in the initial state. However,
this was not clear before the calculations were completed.}
the measurements for $f(x)$ with $10-15$ final steps were then
averaged. Note that these measurements are statistically
correlated. The statistical error for the average result was
estimated by taking the statistical error for the measurement
after $10\ \varepsilon$-steps in the final state (the statistical
error almost does not change from step 10 to 15).

The results are shown in Figs.\ref{kscale}-\ref{ascale}.
First one has to make sure that the longitudinal momentum $K$ was
large enough. In Fig.\ref{kscale} results with $K=\frac{9}{2}$
and $K=\frac{15}{2}$ are compared for some typical choices
of the other parameters. One can easily imagine that
the $K=\frac{9}{2}$ and the $K=\frac{15}{2}$ results lie almost on
the same smooth curve, indicating that $K=\frac{15}{2}$
is large enough. In Fig.\ref{nscale} it is demonstrated how the
results converged as a function of the number of sites.
Note that (with the exception of $N_{sites} = 2$ and $4$, which
were done numerically exactly) although the same number of sweeps
and size of the ensemble were used for all values of $N_{sites}$,
the statistical error bars increase only very slowly with
$N_{sites}$. The same is true for the CPU-time required for the
Monte Carlo calculation. The reason is that the light-cone
vacuum far away from physical particles is trivial.
Thus there are no statistical fluctuations arising from
``updating the vacuum'' on huge lattices (much larger than the
transverse size of the particles).

The $\varepsilon$-dependence is illustrated in Fig.\ref{escale}.
One way to understand Fig.\ref{escale} is to evaluate the
double commutator
which governs the ${\cal O}(\varepsilon^3)$ corrections in
Eq.(\ref{eq:exphab}). The rather lengthy expression will
not be given here, but one can immediately guess the basic
features. First, it is the hopping term
$V_{n,n+1} \propto a^{-2}$ which gives rise to a nonvanishing
commutator in $[H_a,H_b]$, which ``explains'' the increase
of the finite $\varepsilon$ corrections with decreasing $a$:
Fig.\ref{escale} a) vs. b)
(of course, in order to be quantitative one has to
to evaluate the matrix elements which could show an $a$ dependence
as well from the wavefunctions).
Furthermore, the double commutator contains terms
which spread over up to $4$ transverse sites. It is thus
not surprising to find a difference between the finite
$\varepsilon$ effects on 4-site and larger lattices: Fig.\ref{escale}
a) vs. c).
Finally one has to take the $a \rightarrow 0$ limit
(Fig.\ref{ascale}). This is the most difficult part because
one first has to make sure that everything else has
converged for fixed $a$. As discussed above, this requires smaller
$\varepsilon$ for smaller $a$ (hence more steps to project
onto the ground state and hence more computer time).
The required number of lattice sites also increases with
decreasing $a$. Furthermore, since the coupling between sites
goes like $a^{-2}$, the statistical fluctuations induced by this term
increase as $a$ decreases.
An additional reason for the increase of the statistical
fluctuations with decreasing $a$ may lie in the choice of
probabilities (\ref{eq:prob}). On a two site lattice, the
coupling between the two sites is twice as strong as between
adjacent sites on a larger lattice (one step on the lattice
in either direction leads to the same result on a two-site lattice
with periodic boundary conditions). Thus the two site
amplitudes tend to overemphasize highly excited states
(compared to the actual physical situation on a large lattice).
Since this ``driving force'' away from the ground state
increases with smaller $a$, an increase in the statistical
fluctuations results. Evidently, there is still room to
improve the algorithm chosen here. However, since the purpose
of this work is only a feasibility study and since the most
useful choice for the probabilities (\ref{eq:prob}) is expected
to depend on the theory under consideration, this point will not
be elaborated on here any further.

In general, one would extrapolate the results first to
$N_{sites}\rightarrow \infty$ and $\varepsilon \rightarrow 0$
and then analyze the $a$-dependence of the extrapolated results.
Here this is not necessary because the above results demonstrate
that $aN_{sites}=4$ and $\varepsilon = 0.075$ are already
close enough to the continuum limit for
$a \ge 0.025$. Thus one can directly use these results without
having to extrapolate.

While the structure function still changes significantly
as one goes from $(a,N_{sites})=(1,4)$ to $(a,N_{sites})=(0.5,8)$,
there is only a slight difference (at very small $x$ and the slight
``shoulder'' around $x\approx 0.6$)
between
$(a,N_{sites})=(0.5,8)$ and $(a,N_{sites})=(0.25,16)$, i.e. for
$a\le 0.5$ the numerical results (Fig.\ref{ascale}) are
almost independent of $a$ which shows
that the small $a$ scaling region has been reached.
Notice that the wave function renormalization
in $\phi^4_{2+1}$ is finite and thus structure functions
scale to a finite limit as $Q^2 \rightarrow \infty$
(corresponding to $a\rightarrow 0$); i.e. there is
no logarithmic evolution for $\phi^4_{2+1}$ (if there were
logarithmic evolution then the probability to find the state
in the valence configuration would tend to zero as $Q^2 \rightarrow
\infty$).
The shape of these structure functions can be understood
as follows. With lowest order perturbation theory
(fragmentation: valence state $\rightarrow$ 3 partons) one
obtains a structure function which has a maximum near
$x=1/3$.
Once one includes the (nonperturbative)
interactions within the three particle sector,
the structure function becomes smeared out.
The rise at very small $x$ can only be understood
from multiple fragmentation processes. The ``shoulder'' near
$x \approx 0.5-0.6$ is a nonperturbative effect.
It arises because the repulsive $\phi^4$ interaction
tends to enhance components of the wavefunction with a
node in the longitudinal direction. This also explains
why the shoulder is absent for lattices with $4$ or less sites.
The reason is that parity and Bose symmetry require that
a node in the longitudinal direction is accompanied by a node
in the transverse direction. On a small lattice, components of the
wavefunction with a node in the transverse direction
have a very large kinetic energy and are thus suppressed.
The $N_{sites}$-dependence of the results (Fig.\ref{nscale})
indicates that the physical states have a transverse extension
of about $1-2$ in the above units
(i.e. $4-8$ lattice spacings for $a=0.25$)
 because for lattices with
a larger physical volumes, there is no significant volume dependence
of physical observables. However, for more detailed information
one would have to measure transverse density density
correlation functions.

\section{Summary and Conclusions}
I have shown that it is perfectly feasible to perform a Monte Carlo
calculation for a light-front Hamiltonian formulated as a
DLCQ problem on a transverse lattice. The transverse lattice
was used to separate longitudinal and transverse dynamics.
The dynamics within the longitudinal ``sheets'' attached to each
transverse lattice point was solved using DLCQ and conventional
matrix diagonalization algorithms.
The transverse dynamics
was then included by means of Monte Carlo techniques. For this
purpose, the light-cone imaginary time evolution
operator
$<\psi_f | \exp (- N\varepsilon P^-) |\psi_i>$ was approximated
by breaking up the light-front Hamiltonian $P^-$ into
two terms ($P^- = P^-_a + P^-_b$),
each of which contains only interactions between
pairs of sites. $\exp (-N\varepsilon P^-)$
is then evaluated by alternating application of
infinitesimal ``evolution''-operators generated by
$P^-_{a/b}$ respectively. For calculating
the actual path-integral, a variation of
the ensemble projector Monte Carlo
technique was used.\\
In this whole program it was crucial that the longitudinal momentum
$P^+$ was conserved, otherwise the light-front Hamiltonian $P^-$
has no minimum corresponding to a physically meaningful particle
solution. This was the major reason to use DLCQ to solve the
longitudinal dynamics
and not, for example, a longitudinal lattice.
It was furthermore crucial that the
transverse lattice action is local, i.e. it involves only
interactions between neighboring sites. On the one hand, due
to the locality, it was thus possible to perform the abovementioned
breakup of the light front Hamiltonian
$P^- = P^-_a + P^-_b$ in such a way that $P^-_{a/b}$ each can be
written as direct sums of Hamiltonians acting on 2-sites-lattices.
Thus $P^-_{a/b}$ can be easily diagonalized and exponentiated
numerically. The locality of the transverse dynamics was
furthermore important when updating the states at each Monte Carlo
step. This is because locality of the interaction
allowed to formulate the updating of the states at each
$\varepsilon$-step as a sequence of independent local updatings
(this will also be important when running the algorithm on
parallel computers).

The advantages of the  algorithm introduced in this paper
are as follows: most importantly, structure functions are
diagonal in the DLCQ basis used here and are thus easy to
evaluate numerically. Furthermore, since the light-front
momentum $P^+$ is manifestly conserved, and since the
vacuum has $P^+=0$, physical particle states are always manifestly
orthogonal to the vacuum. Thus the Monte Carlo procedure
will always converge to the particle solution with lowest
invariant mass for that particular value of $P^+$
(and the same discrete quantum numbers as the initial state).
In addition,
since the light-front vacuum is trivial, no computer time is
``wasted'' to solve for the vacuum surrounding a physical particle
while one is interested in the particle only (an annoyance for
very large euclidean lattices). Another advantage of using
the light-front Hamiltonian in the Monte Carlo procedure is that
excited states are suppressed by the square of their masses:
$\exp (-N\varepsilon P^-) = \left.\sum_n |n>\exp(-N\varepsilon M_n^2/2P^+)
<n|\right.$
(instead of $\exp( -\beta P^0 ) = \sum_n |n>\exp (-\beta M_n )<n|$
which one encounters in a conventional Hamiltonian formulation).
Thus fewer steps are necessary to filter out the ground state.
The transverse lattice formulation also avoids part of the
species doubling problem for fermions because doublers occur
only for the latticized transverse directions. Thus at most
four species of fermions are generated if one starts from the
naive fermion action in $3+1$ dimensions.
Hence, by means of staggering, one can easily get to
two light flavors of fermions \cite{pauldoubl}.

One of the main disadvantages of the new method are the
occurrence of a few negative matrix-elements in
the infinitesimal evolution operator (although its eigenvalues
are of course still positive). Thus negative scores resulted occasionally
which slightly increased the fluctuations of the signal.
It is expected that this problem gets
worse once fermions are introduced because of the
minus sign in exchange terms (of course this then is nothing
but the usual sign problem for fermions). Another disadvantage
is the explicit breaking of the symmetry between longitudinal
and transverse directions which makes it difficult to recover
full Lorentz invariance in the continuum limit. Although this was no problem in
$\phi^4_{2+1}$, it will be difficult to obtain
Lorentz invariant physical results in general (for theories
where the fundamental particles carry spin).

Numerous extensions of this work are conceivable. First of all,
there is no profound difficulty to extend the
formalism from $2+1$ to $3+1$ dimensions.
Of course, in $3+1$ dimensions $P^- = P^-_a + P^-_b$ has to be
decomposed
in such a way that $P^-_{a/b}$ each can be written as direct
sums of Hamiltonians on 1-plaquette lattices.
Another difference is the dependence of the numerical
results on the lattice spacing $a$:
since $\phi^4_{2+1}$ has a finite wave function renormalization,
scaling is exact which implies that the structure functions
approach a finite limit for $a\rightarrow 0$. Of course in a
renormalizable theory (like $QCD$) this will not be the
case and the structure functions will diverge as $a\rightarrow 0$
--- corresponding to the logarithmic evolution in $Q^2$.
However, at least in principle, this is not a problem
because one can always perform the Monte Carlo calculations
with smaller and smaller spacing $a$ until one can match on
to the perturbative evolution. Since scaling in $QCD_{3+1}$
sets in at moderate $Q^2$ values already, there is reason to
expect that this is also possible in practice (i.e. numerically
practical).

In this work I have investigated only the projector and the
ensemble projector Monte Carlo method because these are
rather straightforward to apply to DLCQ problems.
For more complicated field theories, it may be necessary
to use more efficient techniques, like
guided random walks \cite{epmc2} in the Monte Carlo
procedure.
One could also imagine combining the Monte Carlo technique
presented in this work with renormalization group techniques.
On the one hand this means using perturbative renormalization group
arguments to facilitate the determination of the effective
coupling constants in the light-front Hamiltonian.
On the other hand
one could use the Monte Carlo procedure to perform
nonperturbative studies of renormalization group flow for
light-front Hamiltonians without having to resort to
uncontrolled truncations of the Fock space.
Similarly, it is conceivable that Monte Carlo results
are helpful in determining the effective coupling constants for
the Tamm-Dancoff approach to solving light-front
field theories.
Besides structure functions, one can also use LFEPMC to
calculate valence wavefunctions which have many interesting
applications to various exclusive hard scattering processes
\cite{brolep}.
However, the main question is
whether one can apply the light-front ensemble projector
Monte Carlo to $QCD_{3+1}$.
Here the main difficulty
which remains
is formulating compact (to render the transverse lattice action
gauge invariant) $QCD$ on a transverse lattice using DLCQ or
to construct another approximation to $QCD$ on a transverse
lattice which is suitable for DLCQ.
Once one knows the DLCQ Hamiltonian for $QCD$,
it is straightforward to apply the Monte Carlo technique presented
in this work.
\acknowledgements
This work was supported by the Department of Energy under Grant No.
DE-FG06-90ER40561. The author wishes to thank Michael Frank for
reading and criticizing the manuscript.

\begin{figure}
\caption{Divergent self-energy diagram in $\phi^4_{2+1}$
}
\label{oyster}
\end{figure}

\begin{figure}
\caption{
Result of structure function measurements
for $16$ transverse sites, a transverse spacing of $a=0.5$
and a longitudinal momentum of $K=15/2$ as a function of
the number of $\varepsilon$-steps before taking the overlap with
$\psi_f$. Because of the smaller value of $\varepsilon$ in b),
more $\varepsilon$-steps are necessary to project on the
ground state.}
\label{plateau}
\end{figure}

\begin{figure}
\caption{
Typical examples for structure function measurements
to illustrate the (in)dependence of the results
$K=9/2$ and $K=15/2$.
}
\label{kscale}
\end{figure}

\begin{figure}
\caption{Dependence of the structure function on the
number of transverse sites for the smallest lattice
spacing ($a=0.25$) and the smallest $\varepsilon$-step
used in this work. The longitudinal momentum is $K=15/2$.
Both plots correspond to the same structure function but
at different x-values. In order to avoid overlapping symbols
from different x-values, the results are displayed in two
plots.
}

\label{nscale}
\end{figure}

\begin{figure}
\caption{
$\varepsilon$-dependence of the structure functions for
$K=15/2$. a) for 4 sites and $a=0.5$; b) same as a) but for
$a=0.25$; c) same as a) but for 8 sites.
Note the statistically significant deviation of the results for
$\varepsilon=0.3$ in c) in the intermediate x region, while
a) (same $a$ as c) shows no such deviations.
}

\label{escale}
\end{figure}

\begin{figure}
\caption{
$a$-dependence of the structure functions for $\varepsilon=0.075$
and $K=15/2$. Note that $aN_{sites}$ --- the physical volume ---
is kept fixed in a), b) and c).
}
\label{ascale}
\end{figure}


\begin{references}
\bibitem{bardeen} W. A. Bardeen and R. B. Pearson, Phys.\ Rev.\
D\ {\bf 14}, 547 (1976); W. A. Bardeen, R. B. Pearson and
E. Rabinovici, {\it ibid} {\bf 21}, 1037 (1980).
\bibitem{paul} P. A. Griffin, Mod.\ Phys.\ Lett.\ A\ {\bf 7}, 601
(1992); P. A. Griffin, Nucl.\ Phys.\ B\ {\bf 372}, 270 (1992).
\bibitem{pauldoubl} P. A. Griffin, Phys.\ Rev.\ D\
{\bf 47}, 1530 (1993).
\bibitem{dlcq1} H. C. Pauli and S. J. Brodsky, Phys.\ Rev.\
D\ {\bf 32}, 1993 (1985); {\bf 32}, 2001 (1985); T. Eller,
H. C. Pauli and S. J. Brodsky, Phys.\ Rev.\ D\ {\bf 35}, 1493 (1987).
\bibitem{dlcq2} K. J. Hornbostel, Phys.\ Rev.\ D\ {\bf 45},
3781 (1992).
\bibitem{deu} M. Burkardt, Nucl.\ Phys.\ A\ {\bf 504}, 762 (1990).
\bibitem{lftd} R. J. Perry, A. Harindranath and K. G. Wilson,
Phys.\ Rev.\ Lett.\ {\bf 65}, 2959 (1990); R. J. Perry and
A. Harindranath, Phys.\ Rev.\ D\ {\bf 43}, 4051 (1991).
\bibitem{hari} A. Harindranath and J. Vary, Phys.\ Rev.\ D\ {\bf 36},
1141 (1987).
\bibitem{mbsg} M. Burkardt, Phys.\ Rev.\ {\bf D\ 47}, 4628 (1993).
\bibitem{paulsg} P. A. Griffin, Phys.\ Rev.\ D\ {\bf 46}, 3538
 (1992).
\bibitem{pmc1} J. E. Hirsch, R. L. Sugar, D. J. Scalapino and
R. Blankenbecler, Phys.\ Rev.\ {\bf B\ 26}, 5033 (1982);
J. Potvin and T. A. DeGrand, Phys.\ Rev.\ D\ {\bf 30},
1285 (1984).
\bibitem{epmc1} T. A. DeGrand and J. Potvin, Phys.\ Rev.\
D\ {\bf 31}, 871 (1985).
\bibitem{epmc2} J. W. Negele and H. Orland, ``Quantum Many-Particle
Systems'', (Addison Wesley, Redwood City, 1987).
\bibitem{brolep} S. J. Brodsky and G. P. Lepage, Phys.\ Rev.\
D\ {\bf 22}, 2157 (1980).
\end{references}
\end{document}